\documentstyle[epsfig,12pt,preprint,tighten,aps]{revtex}

\begin{document}

\title{ \rightline{March 2000}
\rightline{Revised April 2000}
\rightline{To appear in Phys.Lett.B}
\title{.}
\title{.}
\vskip 2cm
\ \\
Maximal $\nu_e$ oscillations, Borexino and smoking guns...}

\author{R. Foot}

\address{School of Physics\\
Research Centre for High Energy Physics\\
The University of Melbourne\\
Parkville 3052 Australia\\
foot@physics.unimelb.edu.au}

\maketitle

\begin{abstract}
We examine the maximal $\nu_e \to \nu_s$ and
$\nu_e \to \nu_{\mu,\tau}$ oscillation solutions to the solar
neutrino problem. 
These solutions lead to roughly a $50\%$ solar flux
reduction for the large parameter range
$3 \times 10^{-10} \stackrel{<}{\sim} \delta m^2/eV^2
\stackrel{<}{\sim} 10^{-3}$.
It is known that the earth regeneration effect may cause
a potentially large night-day asymmetry even
for maximal neutrino oscillations. We investigate
the night-day asymmetry predictions 
for the forthcoming Borexino measurement of the 
$^7Be$ neutrinos for both maximal $\nu_e \to \nu_s$
and $\nu_e \to \nu_{\mu,\tau}$ oscillations.
If $y\times 10^{-8}
\stackrel{<}{\sim} \delta m^2/eV^2 \stackrel{<}{\sim}
4y\times 10^{-5}$ (with $y\simeq 0.5$ for $\nu_e \to \nu_s$ case
and $y \simeq 1$ for the $\nu_e \to \nu_{\mu,\tau}$ case)
then the maximal neutrino oscillations 
will lead to observable night-day
asymmetries in Borexino and/or superKamiokande.
With Kamland covering the high mass range,
$10^{-5} \stackrel{<}{\sim} \delta m^2/eV^2
\stackrel{<}{\sim} 10^{-3}$ 
and Borexino/SuperK covering the low mass range,
$3 \times 10^{-10} \stackrel{<}{\sim} \delta m^2/eV^2
\stackrel{<}{\sim} 5\times 10^{-9}$ (``just so'' region), 
essentially all of the
$\delta m^2$ parameter space will soon be scrutinized.

\end{abstract}

\newpage
Maximal oscillations occupy a special point in parameter space.
Neutral Kaons and B-mesons both oscillate maximally with their antiparticle
partners. Interestingly  there is now
strong evidence from solar\cite{solar} and atmospheric\cite{atmos} neutrino
experiments that electron and muon neutrinos also oscillate
maximally with some as yet unidentified partner.
Identifying these states
is one of the most pressing issues in particle physics.

One possibility is that each of the three known neutrinos oscillates 
maximally with an approximately sterile partner.
This behaviour is expected 
to occur if parity is an unbroken symmetry of nature
\cite{flv,flv2}. 
In this theory, the sterile flavour maximally mixing with the $\nu_e$
is identified with the mirror electron neutrino. 
The characteristic maximal mixing feature occurs because 
of the underlying exact parity symmetry between the ordinary 
and mirror sectors. The maximal mixing 
observed for atmospheric muon neutrinos
is nicely in accord with this framework (see e.g.\cite{fvy}), 
which has the atmospheric neutrino problem resolved 
through `$\nu_\mu \to$ mirror partner' oscillations.
Alternatively, it has also been suggested\cite{pd} that
each of the known neutrinos are pseudo-Dirac
fermions\cite{pd2} which has each of the known neutrinos
oscillating maximally into a sterile, $\nu_R$ partner. 
Both of these ideas motivate the study of maximal
two flavour $\nu_e \to \nu_s$ oscillations (where $\nu_s$ means sterile
neutrino).

Of course there are other possibilities. For example it is
possible that the neutrino anomalies
are due to bi-maximal mixing\cite{bmax}.
This sees the atmospheric anomaly being solved by
maximal $\nu_\mu \to \nu_{\tau}$ oscillations and
the solar problem being solved by maximal
$\nu_e \to (\nu_\mu + \nu_\tau)/\sqrt{2}$ oscillations.
The bi-maximal hypothesis is an interesting possibility even
though a compelling theoretical motivation for it
has yet to be found.
Thus, two flavour maximal $\nu_e \to \nu_{\mu,\tau}$ oscillations
(where $\nu_{\mu,\tau}$ means any linear combination 
of $\nu_\mu$ or $\nu_\tau$) is therefore also interesting.
Note that the two phenomenologically similar 
(but theoretically very different) possibilities
of $\nu_e \to \nu_s$ and $\nu_e \to \nu_{\mu,\tau}$ oscillations
will hopefully be distinguished at the Sudbury Neutrino Observatory
(SNO)\cite{sno} when they
measure the neutral and charged current contributions separately.

Two flavour maximal oscillations between the electron neutrino 
and a sterile or active flavor
produces an approximate $50\%$ 
solar neutrino flux reduction
for a large range of $\delta m^2$:
\begin{equation}
\text{3}\ \times 10^{-10}
\stackrel{<}{\sim} \frac{\delta m^2}{{\rm eV}^2} \stackrel{<}{\sim} 
10^{-3}.
\label{1}
\end{equation}
The reason why the reduction is not exactly $50\%$ is
because earth regeneration effects\cite{bou}
can modify the night time rate (and there is also a small neutral
current contribution in the case of active neutrino oscillations
in $\nu e \to \nu e$ elastic scattering experiments).
This earth regeneration effect can lead to a modest energy 
dependence, but not enough to
explain the low Homestake result.
The upper bound in Eq.(\ref{1}) arises from the lack of 
$\overline \nu_e$ disappearence in the CHOOZ
experiment\cite{chooz}\footnote{Note that this 
entire range for $\delta m^2$ does not necessarily lead 
to any inconsistency with bounds imposed by big bang 
nucleosynthesis\cite{footbb}.}, while the lower bound 
can be deduced from the observed 
recoil electron energy spectrum.
For $E_{recoil} < 12\ MeV$ the
recoil electron energy spectrum is consistent
with an overall flux reduction of roughly $50\%$ with
no evidence of any energy dependent distortion of the
neutrino flux.
Maximal oscillations with $\delta m^2 \stackrel{<}{\sim}
3 \times 10^{-10}\ eV^2$ either significantly distort
this spectrum or (in the case of very small $\delta m^2$)
do not lead to any flux reduction (because the oscillation
length becomes too long for oscillations to have any effect).
Note that there is a hint of a spectral
anomaly for $E_{recoil} > 12\ MeV$\cite{suz} which may be
due to ``just so" oscillations \cite{bar}
with $\delta m^2 \sim 
4\times 10^{-10} \ eV^2$ (see e.g.\cite{oth,cfv}) although it is also
possible
that it is due to a systematic uncertainty or statistical
fluctuation.

The current experimental situation for solar neutrinos is 
summarized in the table below where the data is compared to
the theoretical model of Ref.\cite{bp}.
\vskip 0.4cm

{\begin{center}
\begin{tabular}{|l|l|l|}
\hline
Experiment$\;\;\;\;\;\;\;$
&Flux$\;\;\;$ 
&Theory$\;\;\;\;\;\;\;\;\;\;\;\;$\\
\hline
Homestake\cite{solar}&$ 2.55 \pm 0.25 (stat+syst)$ 
SNU&$7.7^{+1.2}_{-1.0} $ SNU\\
Kamiokande\cite{solar}&$2.80\pm 0.19(stat)\pm 0.33 (syst)
\times 10^6 cm^{-2}s^{-1}$&$5.15^{+1.0}_{-0.7}$ $10^6 cm^{-2}s^{-1}$\\
SuperKamiokande\cite{solar}&$2.44\pm 0.05(stat)\pm 0.08 (syst)
\times 10^6 cm^{-2}s^{-1}$&$" \  \  \  " \  \    \   "$\\
GALLEX\cite{solar}&$77 \pm 6 (stat) \pm 5 (syst)$ SNU&
$129^{+8}_{-6}$ SNU\\
SAGE\cite{solar}&$67 \pm 7 (stat) \pm 3.5 (syst)$ SNU&
$" \  \  \ " \   \    \  "$ \\
\hline
\end{tabular}\end{center}}

\noindent
Table Caption:
Comparison of solar neutrino experiments with 
the solar model of Ref.\cite{bp}. 

\vskip 0.4cm

\noindent
As the above table shows,
the approximate $50\%$ flux reduction implied by
maximal neutrino oscillations in the parameter range,
Eq.(\ref{1}) would reconcile 
four out of the five experiments which means that this solution
is in broad agreement with the experiments. 
The misbehaving experiment is Homestake which is roughly
$3-4$ standard deviations too low (a $50\%$ flux reduction would
imply $\sim 3.3-4.5\ SNU$ c.f. the measured $2.55\pm 0.25\
SNU$).  If taken seriously, then
the low Homestake results suggests some
specific regions of parameter space\cite{sm}. However
one should keep in mind that theoretical solar models involve 
a number of simplifying assumptions and it is therefore
also possible that the $^7Be$ neutrino flux has been overestimated
which would alleviate the discrepancy.
Alternatively, there might be some
as yet unidentified systematic error in the Homestake experiment.
This seems plausible as the Homestake team argued that
their data was anti-correlated with the sun spot cycle during
the period before about 1986 (with high confidence level), but has 
since stabilized (see e.g. Ref.\cite{mor} and also section 10.5
of Ref.\cite{bah} for some discussion about this).
We adopt the cautious viewpoint that
this experiment needs to be checked by another experiment
before a compelling case for large energy {\it dependent} suppression
of the solar flux can be made.

Recently, Guth et al\cite{guth} pointed out that the 
earth regeneration effect\cite{bou} leads to a night-day
asymmetry, $A_{n-d}$, for maximal neutrino oscillations. 
We define $A_{n-d}$ by\footnote{
Note that in the literature an alternative definition is also
used which differs from
our definition in Eq.(\ref{ess}) by an approximate factor of 2.}
\begin{equation}
A_{n-d} \equiv {N - D \over N + D}.
\label{ess}
\end{equation}
Guth et al computed the night-day asymmetry for
superKamiokande for large angle and maximal $\nu_e \to \nu_{\mu,\tau}$
oscillations. In Ref.\cite{cfv} this was extended to 
maximal $\nu_e \to \nu_s$ oscillations where it was shown
that the current measurements of the night-day asymmetry
allow the parameter space $2\times 10^{-7}\stackrel{<}{\sim}
\delta m^2/eV^2 \stackrel{<}{\sim} 8\times 10^{-6}$ to be excluded
at about two standard deviations.
The point of this
paper is to study both maximal $\nu_e \to \nu_s$ and
$\nu_e \to \nu_{\mu,\tau}$ oscillation solutions in the context
of the forthcoming Borexino experiment.

The Borexino experiment\cite{bor} is a real time $\nu  e \to \nu e$ 
elastic scattering experiment
like superKamiokande, but is designed to be sensitive to relatively
low energy neutrinos. This should allow the neutrino flux from
the $E = 0.86\ MeV$ $^7Be$ line to be measured. 
Our procedure for calculating the night-day asymmetry is
very similar to Refs.\cite{guth,cfv} so we will not 
repeat the details here. One difference is 
that now we must use the zenith distribution function for the 
Gran Sasso latitude which we obtain from Ref.\cite{jnb}. 
Also, we use the 
advertised\cite{bor,gbr} Borexino cuts in the apparent recoil 
electron kinetic energy of
$0.25 < E_{recoil}/MeV < 0.70$. With this cut, about
$80\%$ of the recoil electron events are due to 
$^7Be$ neutrinos and $20\%$ due to CNO and pep neutrinos\cite{gbr}.

Our results for the night-day asymmetry for the maximal
$\nu_e \to \nu_s$ oscillation solution are given in figure 1
(solid line)
and the maximal $\nu_e \to \nu_{\mu,\tau}$ oscillation solution
is given in figure 2. Also shown (dashed line) is 
the analogous results obtained for 
the superKamiokande experiment obtained from Ref.\cite{cfv}.
Also included (dotted line) in the figures is the results for Kamland
which may also be able to measure low energy solar
neutrinos\cite{kam}.

As far as I am aware,
the night-day asymmetry for $\nu_e \to \nu_s$ oscillations (maximal
or otherwise) has never been 
computed previously in the context of Borexino.
While this paper was in preparation we became aware
of the recent eprint, Ref.\cite{gfm} which discusses
the night-day asymmetry for 
large angle $\nu_e \to \nu_{\mu,\tau}$ oscillations in the
context of Borexino.
Our results are in agreement with the results of
this paper when we examine the $sin^2 2\theta = 1$
line on their
contour plot in the $\delta m^2, \sin^2 2\theta$ plane. 
For the subset of people interested mainly in maximal
mixing our results are complementary to those of
Ref.\cite{gfm} since they contain more information than
the contour plots. 

The night-day asymmetry results for Borexino are roughly similar to the
results for superKamiokande, except they are shifted to lower
values of $\delta m^2$. This shift of about an order of magnitude
in $\delta m^2$ is quite easy to understand. It arises because 
the typical neutrino energies  
for superKamiokande are about an order of magnitude larger
than the energies relevant for Borexino and 
the oscillations depend on $E, \delta m^2$ only in the
ratio $E/\delta m^2$.

Assuming maximal oscillations in the
range, Eq.(\ref{1})
(and the solar model of Ref.\cite{bp}),
Borexino is 
expected\cite{gbr} to detect around 25-30 events/day
(with the cut $0.25 < E_{recoil}/MeV < 0.70$).
This is somewhat more than in the SuperKamiokande experiment.
Accordingly a night-day asymmetry as low as
$A_{n-d} \sim 0.02$ (or even lower) maybe observable at Borexino after
only a couple of years of data (see Ref.\cite{c14,gfm} 
for discussions of backgrounds and systematic
uncertainties). 
From our figures we see that the maximal neutrino oscillation
solutions lead to a significant
(i.e. $A_{n-d} \stackrel{>}{\sim} 0.02$)
night-day asymmetry in Borexino 
and/or superKamiokande for the parameter range:
\begin{eqnarray}
5\times 10^{-9} \stackrel{<}{\sim} \delta m^2/eV^2 
\stackrel{<}{\sim} 2\times 10^{-5} \ for \ \nu_e \to \nu_s \nonumber \\
10^{-8} \stackrel{<}{\sim} \delta m^2/eV^2 
\stackrel{<}{\sim} 4\times 10^{-5} \ for \ \nu_e \to \nu_{\mu,\tau}
\label{med}
\end{eqnarray}
If $\delta m^2$ is in this range then the night-day asymmetry
should provide a suitable ``smoking gun"  signature which could provide
compelling evidence that the solar neutrino problem is solved
by neutrino oscillations. 
This is especially important for
$\nu_e \to \nu_s$ oscillations since it predicts that SNO will
not find any anomalous NC/CC ratio.

Let us label the region in Eq.(\ref{med}) as
the ``medium $\delta m^2$ region". Observe that there are two other
possible regions of interest: The ``high $\delta m^2$ region"
with $2\times 10^{-5} \stackrel{<}{\sim} \delta m^2/eV^2 
\stackrel{<}{\sim} 10^{-3}$ and
the ``low $\delta m^2$ region" with
$3 \times 10^{-10} \stackrel{<}{\sim} \delta m^2/eV^2 \stackrel{<}{\sim} 
5\times 10^{-9}$ (where the upper boundary is increased
to about $10^{-8}$ for $\nu_e \to \nu_{\mu,\tau}$ oscillations).
If $\delta m^2$ is in the high region then
the Kamland experiment will 
be able to see reactor electron neutrino disappearance. 
This should fully test this region.
Note that part of the high $\delta m^2$ region
is already being probed by 
the atmospheric neutrino experiments.
For large values of $\delta m^2 \stackrel{>}{\sim}
10^{-4} eV^2$, $\nu_e \to \nu_s$  
oscillations lead
to observable up-down asymmetries for the detected electrons\cite{bfv}.
At the moment there is no evidence for any electron up-down asymmetry
which disfavours maximal $\nu_e \to \nu_s$ oscillations
with $\delta m^2/eV^2 \stackrel{>}{\sim} 10^{-4}$ 
(similar results
should also hold for $\nu_e \to \nu_{\mu,\tau}$ oscillations).
For $\delta m^2$ in the low region the oscillations
will lead to ``just so" phenomena
such as energy distortion and seasonal effects.
These effects can be probed at superKamiokande for $\delta m^2/eV^2
\stackrel{<}{\sim} 10^{-9}$ (see e.g.\cite{oth,cfv}) 
and at Borexino for $\delta m^2/eV^2 \stackrel{<}{\sim}
5\times 10^{-9}$\cite{mur}.

We summarize the current situation and expected 
sensitivities to $\delta m^2$ of the various experiments
in figure 3 (for the maximal $\nu_e \to \nu_s$ oscillations)
and figure 4 (for the maximal $\nu_e\to \nu_{\mu,\tau}$ oscillations).
In the $\nu_e \to \nu_s$ case observe that all of
the $\delta m^2$ parameter space will lead to
a ``smoking gun'' signature
in at least one of the experiments (Borexino, SuperKamiokande and/or
Kamland).
For the maximal $\nu_e \to \nu_{\mu,\tau}$
oscillations, there is a narrow region 
$5\times 10^{-9} \stackrel{<}{\sim} \delta m^2 
\stackrel{<}{\sim} 10^{-8}$ which may fall between the
cracks. This region may possibly be tested at 
Borexino (or Kamland) if their systematic uncertainties can be
reduced sufficiently so that $A_{n-d} \sim 0.01$
(cf.Ref.\cite{gfm})
could be seen for the $^7Be$ neutrinos.

Finally, the current superKamiokande
measurement of the night-day asymmetry is\cite{hay}
\begin{equation}
A_{n-d} = 0.033\pm 0.017 \ (stat+syst).
\end{equation}
If we take the above hint seriously, i.e. that the superKamiokande
night-day asymmetry is small but non-zero
then in the context of the maximal mixing scenario
there are two possible regions for $\delta m^2$, depending
on which side of the night-day ``mountain'' we are on.
If we are on the left-hand slope then Borexino will see
a large night-day asymmetry. Our results in figure 1,2
suggest a range of 
$0.12 < A_{n-d} < 0.20$ for the $\nu_e \to \nu_s$ case
and $0.10 < A_{n-d} < 0.16$ for the $\nu_e \to \nu_{\mu,\tau}$
case.
Of course if we are on the right-hand slope of the superKamiokande 
night-day mountain then Borexino will not see any night-day asymmetry.
The shape of the superKamiokande energy spectrum of the night-time events
can also tell us, in principle, which side of the night-day 
mountain we are on (see e.g.\cite{guth,cfv}).
 
In summary, there are strong general and specific theoretical reasons
for neutrino oscillations to be maximal.
This prejudice
is broadly consistent with the $\nu_\mu$ disappearance observed
by the atmospheric neutrino experiments as well as the
$\nu_e$ disappearance suggested by the solar neutrino experiments.
We have examined the predictions of maximal $\nu_e$ oscillations
for Borexino (see figures 1,2).
This experiment together with SNO, superKamiokande and
Kamland 
should be able to cover essentially all of the parameter space
of interest.

\vskip 0.3cm
\noindent
{\bf Acknowledgements}
\vskip 0.2cm
\noindent
The author would like to thank Silvia Bonetti and Marco Giammarchi 
for answering my questions about the sensitivity
of Borexino to pp neutrinos. The author would also like
to thank H. Murayama for some comments which lead to 
improvements to the paper. 

{\bf Figure Captions}
\vskip 0.7cm
\noindent
Figure 1: Night-day asymmetry, $A_{n-d}\equiv (N-D)/(N+D)$ versus 
$\delta m^2/\text{eV}^2$ for maximal $\nu_e \to \nu_s$ oscillations.
The solid line is the prediction for Borexino assuming a cut on the 
apparent recoil electron energy of $0.25 < E_{recoil}/MeV < 0.70$, while
the dashed line is the night-day asymmetry for superKamiokande
($6.5 < E_{recoil}/MeV < 20$).
Also shown (dotted line) is the corresponding result for the 
Kamland site ($0.25 < E_{recoil}/MeV < 0.70$).

\vskip 0.5cm
\noindent
Figure 2: Same as figure 1 except for maximal $\nu_e \to
\nu_{\mu,\tau}$ oscillations.

\vskip 0.5cm
\noindent
Figure 3: Sensitivity of maximal $\nu_e \to \nu_s$ oscillations
to the various experiments.
Note that the ``SuperK night-day" region denotes the region
with an observable ($A_{n-d} \stackrel{>}{\sim} 0.02$)
night-day asymmetry at superKamiokande (which is not so large
as to be excluded by the current superKamiokande data).

\vskip 0.5cm
\noindent
Figure 4: 
Same as figure 3 except for maximal $\nu_e \to \nu_{\mu,\tau}$ oscillations.

\newpage
\epsfig{file=bfig1.eps,width=15cm}
\newpage
\epsfig{file=bfig2.eps,width=15cm}
\newpage
\epsfig{file=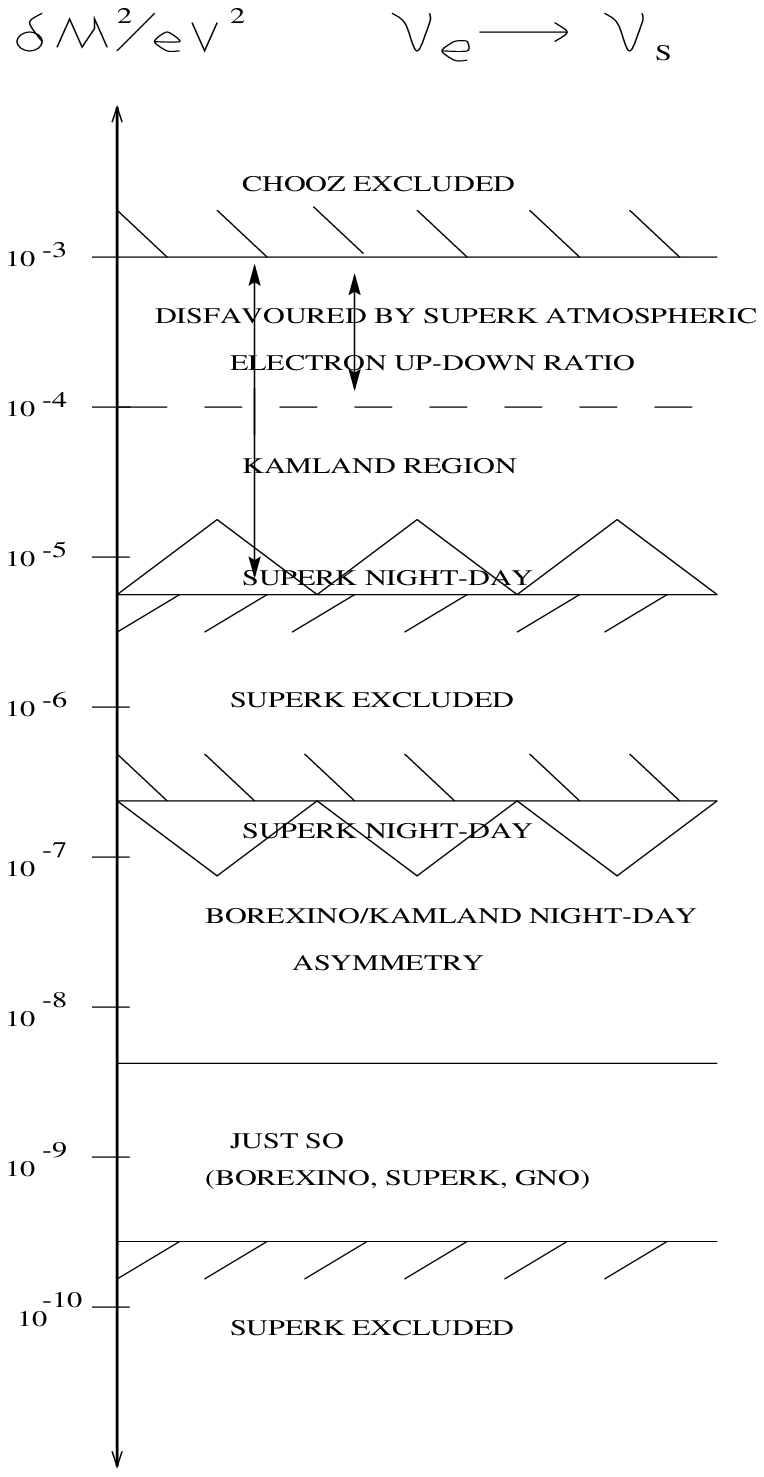,width=15cm}
\newpage
\epsfig{file=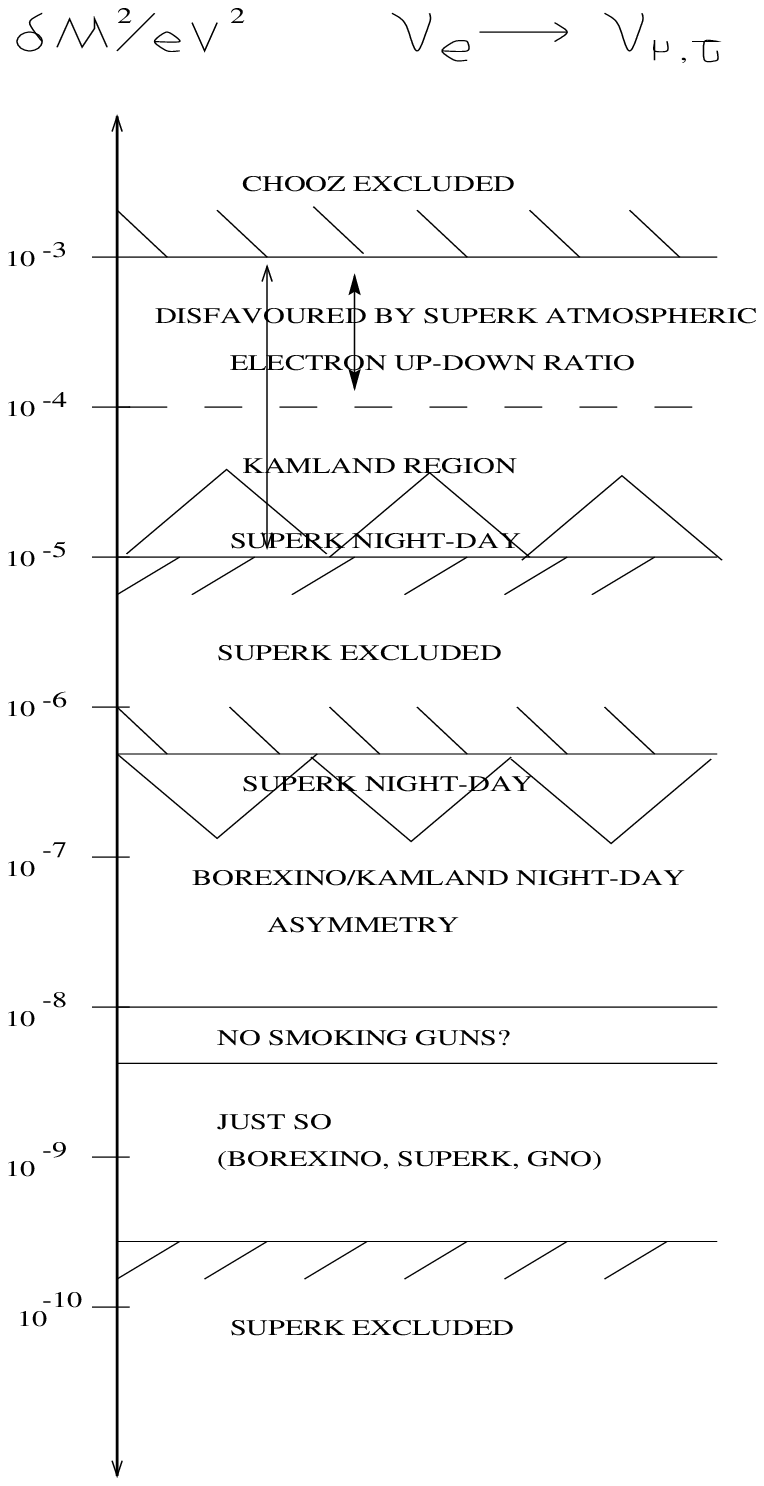,width=15cm}
\end{document}